# Spin-wave-driven skyrmion dynamics in ferrimagnets: Effect of net angular momentum

Y. Liu[1], T. T. Liu[1], Z. Jin[1], Z. P. Hou[1], D. Y. Chen[1], Z. Fan[1], M. Zeng[1], X. B. Lu[1], X. S. Gao[1], M. H. Qin[1,*], and J. –M. Liu[1,2]

[1]*Guangdong Provincial Key Laboratory of Quantum Engineering and Quantum Materials and Institute for Advanced Materials, South China Academy of Advanced Optoelectronics, South China Normal University, Guangzhou 510006, China*

[2]*Laboratory of Solid State Microstructures, Nanjing University, Nanjing 210093, China*

**[Abstract]** Searching for low-power-consuming and high-efficient methods for well controllable driving of skyrmion motion is one of the most concerned issues for future spintronic applications, raising high concern with an appreciated choice of magnetic media and driving scenario. In this work, we propose a novel scenario of spin wave driven skyrmion motion in a ferrimagnetic (FiM) lattice with the net angular momentum $\delta_s$. We investigate theoretically the effect of both $\delta_s$ and the circular polarization of spin wave on the skyrmion dynamics. It is revealed that the momentum onto the skyrmion imposed by the excited spin wave can be partitioned into a ferromagnetic term plus an antiferromagnetic term. The ratio of these two terms and consequently the Hall angle of skyrmion motion can be formulated as the functions of $\delta_s$, demonstrating the key role of $\delta_s$ as an effective control-parameter for the skyrmion motion. Moreover, the spin wave frequency dependent skyrmion motion is discussed, predicting the frequency enhanced skyrmion Hall motion. This work thus represents an essential contribution to understand the skyrmion dynamics in a FiM lattice.

[*]Email: qinmh@scnu.edu.cn

# I. Introduction

Magnetic skyrmions are particle-like spin textures which are often observed in chiral magnets with Dzyaloshinskii-Moriya interaction (DMI) due to the broken inversion symmetry. Owing to the topologically protected property, skyrmions are rather stable with nanoscale in size [1–4], providing them possibility to be information carrier in future spintronic devices. Recently, a number of theoretical and experimental studies on skyrmions creation and annihilation [5–8], dynamical manipulation [2,9–11], and fabrication of skyrmion-based devices [12–15] have been reported. For driving the motion of skyrmion, various technical scenarios are proposed, including those utilizing electric current [10,16,17], magnetic field [18–20], and magnetic anisotropy gradient [21,22] etc. While these scenarios show their respective advantages, shortcomings are also concerned. For example, spin-polarized electric current driving always generates Joule heat that is unfavorable. The implementation of magnetic anisotropy gradient is somewhat limited due to the induced complexity in device integration. The low-power-consuming and high-efficient control scenario for skyrmion motion remains to be one of the most essential issues for potential application of skyrmion-spintronics.

For ultra-low power consumption, one promising choice is utilization of spin wave or so-called magnon [23–26] that can drive skyrmion motion without Joule heating and thus is highly preferred [13,27,28]. Moreover, unlike electric current driving, spin wave as driving source is free of physical transport of charges and thus applicable in insulating systems, an unbeatable advantage. This motivation has been explored for years and so far the skyrmion motion driven by spin waves in both ferromagnetic (FM) and antiferromagnetic (AFM) media have been investigated. In ferromagnets, by definition spin wave is only right-circularly polarized because of the broken time-reversal symmetry [29]. This property enables the motion of skyrmions towards the spin wave source in the presence of a transverse motion (i.e. the Hall motion), and the magnon-skyrmion interaction could be viewed as an elastic scattering process where the skyrmion behaves like a massless particle [30,31]. However, for antiferromagnets, spin wave can be either left- or right-circularly polarized [13,29,32], which provides the polarization degrees of freedom including all linear and elliptical polarizations for practical utilization. Indeed, earlier works revealed the dependence of skyrmion motion on the spin-wave polarization in antiferromagnets, and specifically the motion is guided towards the left (right) transverse direction, as driven by the excited right- (left-) handed spin waves. Besides the handedness dependence, it was found that this Hall motion also depends on the linear polarization of spin wave, as unveiled in our earlier work [33].

While the spin wave driven motion of skyrmions in FM and AFM systems has been demonstrated, the as-raised technical issues are also discussed. The strong stray field and relatively slow spin dynamics are the disadvantages for a FM system, while effective detection and controllability of an AFM texture are still quite challenging in practice due to the zero net magnetic moment although its high-speed spin dynamics and negligible stray field are highly appreciated. Along this line, a compromise is going to ferrimagnetic (FiM) systems which are under intensive discussion for their preference in future spintronic devices. A FiM lattice has a nonzero net moment even around the angular momentum compensation temperature ($T_A$). While the spin dynamics comparable to antiferromagnets can be reserved if the net angular momentum $\delta_s$ is small, the net moment, generally weak in the vicinity of $T_A$ [33–35], can be used as the sensitive parameter to characterize the magnetic state of a FiM system without inducing large stray field. It is also straightforward to argue that the skyrmion motion and its dynamics in a FiM system would be a generalized extension of the dynamics of skyrmion motions in FM and AFM systems, given $\delta_s = 0$ for an AFM system and sufficiently large $\delta_s$ for a FM system.

Therefore, one may choose a proper and nonzero $\delta_s$ for probing the magnetic state in a spin wave driven skyrmion device made of a FiM system, and thus $\delta_s$ appears to be a core control parameter for such device. For example, earlier work demonstrated that the kinetic energy and inertial energy of the Berry phase for a FiM system depend on parameter $\delta_s$, suggesting a possible modulation of spin dynamics [35–38]. It should be mentioned that parameter $\delta_s$ can be easily adjusted by ion doping or temperature. By this strategy, it is definitely deserved to explore how this parameter is used as an additional degree of freedom in controlling the spin wave driven skyrmion motion. Given a series of emergent phenomena / effects for such motion in FM and AFM systems [39], our major concern comes to these phenomena in a FiM system in response to spin wave. This study thus not only uncovers likely additional phenomena but also provides a generalized scenario for skyrmion motion manipulation and device control. For instance, skyrmion Hall motion which is a highly concerned drawback for applications may be suppressed in a FiM system if a suitable $\delta_s$ value is chosen.

In fact, skyrmion generation and its motion in FiM materials were observed recently [2,40–43], enhancing the priority for exploring the spin wave driven skyrmion motion in a FiM system. In this work, we study the skyrmion motion in a FiM lattice under the stimulation of a circularly polarized spin wave, using rigorous treatment and numerical methods. Net angular momentum $\delta_s$ used as the control parameter of the skyrmion motion and the effect of circular polarization (left- and right-circularly) of the spin wave on the skyrmion motion will be discussed in details. It is revealed that the motion momentum onto a skyrmion, acted by an excited spin wave, may be rationally partitioned into two separate terms. One is equivalent to the term existing in a FM lattice and the other is similar to the term existing in an AFM lattice, an expectable outcome. Moreover, it is demonstrated that $\delta_s$ as the control parameter determines the ratio of the two terms, which in turn controls the skyrmion Hall motion behavior. These rigorous treatments can be further evidenced by the atomistic spin simulations based on solving the two coupled Landau-Lifshitz-Gilbert (LLG) equations, illustrating the dependence of the Hall angle on $\delta_s$.

## II. Theory for skyrmion dynamics driven by circularly polarized spin waves

In this section, we first present a full set of definitions on magnetic structure for a FiM system where two sublattices are usually assumed, and then derive the equations of motion for a skyrmion.

In the continuum framework, the local spin densities are denoted by $\mathbf{s}_1 = s_1 \mathbf{n}_1$ and $\mathbf{s}_2 = s_2 \mathbf{n}_2$, where $\mathbf{n}_1$ and $\mathbf{n}_2$ are unit vectors of each of the two FiM sublattices, $s_i = M_i/\gamma_i$ is the magnitude of the spin density, $M_i$ is the magnetization and $\gamma_i = g_i \mu_B/\hbar$ is the gyromagnetic ratio of sublattice $i$. For clarity, both $s_1$ and $s_2$ are defined to be positive, $s = (s_1 + s_2)/2$ is the spin density average of the two sublattices and thus the net angle momentum $\delta_s = s_1 - s_2$.

Furthermore, one may introduce the staggered vectors $\mathbf{n} = (\mathbf{n}_1 - \mathbf{n}_2)/2$ and $\mathbf{m} = (\mathbf{n}_1 + \mathbf{n}_2)/2$ to deal with the dynamic equations of ferrimagnets. With this set of definitions, one can formulate the dynamic equation of the matter under study.

### A. Lagrangian of FiM dynamics

Following earlier work, the FiM dynamic equations can be written in terms of the Néel

vector **n**, whose Lagrangian density is given by [2,34,44,45]

$$L = L_B - U,  \qquad (1)$$

where $L_B$ is the spin Berry phase and $U$ is the free energy density. The spin Berry phase, which governs the magnetic dynamics, is given by:

$$L_B = \delta_s \mathbf{a}(\mathbf{n}) \cdot \dot{\mathbf{n}} + \rho \dot{\mathbf{n}}^2, \quad \dot{\mathbf{n}} = d\mathbf{n}/dt, \qquad (2)$$

where $\rho = s^2/a$ is the constant of inertia [34] with the homogeneous exchange constant $a$, $\mathbf{a}(\mathbf{n})$ is the vector potential for the assumed magnetic monopole [10,30]. The first term represents the spin Berry phase associated with $\delta_s$, and the second term represents the dynamic inertia of **n**.

The free energy density $U$ including the exchange interaction, interfacial Dzyaloshinskii-Moriya interaction (DMI), and uniaxial anisotropy energy, is defined as:

$$U = A(\nabla \mathbf{n})^2 - K \mathbf{n}_z^2 + D[\mathbf{n}_z \nabla \cdot \mathbf{n} - (\mathbf{n} \cdot \nabla)\mathbf{n}_z], \qquad (3)$$

where $A$, $K$ and $D$ are the approached exchange, anisotropy and interfacial DMI coefficients, respectively.

In proceeding, one considers the excitation of spin wave in such a FiM system. Usually, this excitation is described by the Néel vector **n** that can be divided into two parts [13,24]: (1) the slowly moving texture $\mathbf{n}_0$, (2) the fast evolving spin-wave excitation component **n'**, and they are called respectively the static and dynamic components, and are more conveniently described when local coordinate $\hat{\mathbf{e}}_{\theta,\varphi,r}$ is included [30]. Namely, $\mathbf{n}_0 = \hat{\mathbf{e}}_r$ and $\mathbf{n}' = n_\theta \hat{\mathbf{e}}_\theta + n_\varphi \hat{\mathbf{e}}_\varphi$. A complex expression of this staggered vector reads $\psi(\mathbf{r}, t) = n_\theta + i n_\varphi$. Consequently, the spin Berry phase term $L_B$ of the Lagrangian density $L$ and the free energy density $U$ in Eq. (1) are re-written as:

$$\begin{aligned}L_B &= \delta_s\left(a_t^0 - \phi a_t^0\right) + \rho\left[(1-2\phi)\dot{\mathbf{n}}_0^2 + i\left(\psi^* \partial_t \psi - \psi \partial_t \psi^*\right)a_t^0 + \partial_t \psi^* \partial_t \psi\right]/2 \\ U &= u_0 + \mathbf{j} \cdot \mathbf{a}_{\text{total}} - 2\phi u_0 + A \nabla \psi^* \nabla \psi / 2 - K\phi \end{aligned} \qquad (4)$$

where $\mathbf{j} = iA(\psi^* \nabla \psi - \psi \nabla \psi^*)/2$ is the flux of spin waves and $\phi = \psi^*\psi/2$ is the local intensity of spin waves, $u_0 = u(\mathbf{n}_0)$ is the local texture energy associated with the moving texture $\mathbf{n}_0$ as the background, $\mathbf{a}_{\text{total}} = \mathbf{a}_0 + \mathbf{a}_D$ is the total vector potential, including the inhomogeneous magnetization and DMI, $a_t^0$ is the scalar potential associated with the dynamics of spin texture [30,36].

*B. Collective coordinate theory*

Given the FiM dynamic equations, now one can focus on the skyrmion formation and its motion, as driven by spin wave. Since the issue to be discussed is on the moving objects, and both skyrmion and spin wave can be treated as rigid texture and quasi-particles moving in the lattice, one needs to construct the collective coordinates for the issue to be discussed.

Generally, a skyrmion can be expressed in the collective coordinates $\mathbf{n}_0(t) = \mathbf{n}_0[\{X_\mu(t)\}]$. For the spin wave, the collective coordinates $\psi(t) = \psi[\{x_i(t)\}]$ are usually employed, as done in wave packet theory [44,46]. Here subscript indices ($\mu$, $\nu$) and ($i$, $j$) correspond to the two-dimensional real-space vectors for skyrmion and spin wave packets collective coordinates, respectively. The whole Lagrangian $L$, here specifically denoted as $L_z$, can be now represented in terms of $\{X_\mu(t)\}$ and $\{x_i(t)\}$:

$$L_z = \delta_s A_\mu^0(\phi) \dot{X}_\mu + \rho \left[ M^{\mu\nu} \dot{X}^2 - 4\omega \rho_{sw} a_\mu^0 \dot{X}_\mu + \rho_{sw} \omega^2 \right]/2 \\ - U_0(\phi) - \rho_{sw} \left( \Omega_\pm \dot{\mathbf{x}} \cdot \mathbf{a}_{total} - \Omega_\pm^2 \dot{x}_i^2 / 4A \right) \quad , \tag{5}$$

where $\rho_{sw} = \int \phi dV$ is the total intensity of spin waves, $M^{\mu\nu} = (1 - 2\phi)\int(\partial_\mu \mathbf{n}_0 \cdot \partial_\nu \mathbf{n}_0)dV$ is the effective mass of skyrmion, $A_\mu^0(\phi) = \int(1 - \phi)a_t^0 dV \approx A_\mu^0 - \rho_{sw} \cdot a_\mu^0(\mathbf{x})$ is the vector potential for coordinate $X_\mu$ with $a_\mu^0 = \mathbf{a}(\mathbf{n}_0) \cdot \partial_\mu \mathbf{n}_0$, $U_0(\phi) = \int(1 - 2\phi)u_0 dV \approx U_0 - 2\rho_{sw} u_0(\mathbf{x})$ denotes the total texture energy with $A_\mu^0$ and $U_0$ being the vector potential and energy of the spin texture (no spin wave component), respectively, and $\dot{\mathbf{x}}$ is the velocity vector of spin wave packet. $\Omega_\pm = 2\rho\omega \pm (-\delta_s)$, where $\Omega_+$ ($\Omega_-$) corresponds to the right- (left-) handed spin wave.

Given the Lagrangian formulation, one needs to consider the resistance to the motion of objects, i.e. the friction force against the motion which should be proportional linearly to the motion velocity, described as the Rayleigh term $R$ for the dissipation in the collective coordinates:

$$R = s \cdot \alpha \left[ \Gamma_{\mu\nu} \dot{X}_\mu \dot{X}_\nu + 2\rho_{sw} \kappa \dot{x}_i^2 \right]/2 \\ \Gamma_{\mu\nu} = (1 - 2\phi) \int (\partial_\mu \mathbf{n}_0 \cdot \partial_\nu \mathbf{n}_0) dV \quad . \tag{6} \\ \kappa = \left[ (2\rho\omega - \delta_s)^2 - \delta_s^2 - 4\rho K \right]/4A\rho$$

where $\alpha = (s_1\alpha_1 + s_2\alpha_2)/s$, $s = (s_1 + s_2)/2$.

Subsequently, the motion equations of the skyrmion and spin wave packet in the collective coordinates using the Euler-Lagrangian rule are expressed as:

*motion equation for skyrmion:*

$$\rho M^{\mu\nu}\ddot{X}_\nu = E_\mu^0 + (\delta_s - 2\rho\omega\rho_{sw})B_{\mu\nu}^0 \dot{X}_\nu - s\cdot\alpha\Gamma_{\mu\nu}\dot{X}_\nu - \rho_{sw}\Omega_\pm b_{\mu i}\dot{x} \quad (7a)$$

*motion equation for spin wave packet:*

$$\Omega_\pm^2 m_{sw}\ddot{x}_i = 2e_i - \Omega_\pm b_{ij}\dot{x}_j - 2s\cdot\alpha\kappa\dot{x}_i + (2\rho\omega - \delta_s)b_{\mu i}\dot{X}_\mu \quad (7b)$$

where $E_\mu^0 = -\partial_\mu U_0$, $e_i = -\partial_i u_0$, $B_{\mu\nu}^0 = \partial_\mu A_\nu^0(\phi) - \partial_\nu A_\mu^0(\phi)$, $b_{\mu i} = \partial_\mu a_i - \partial_i a_\mu$, $b_{ij} = \partial_i a_j - \partial_j a_i$ are the effective electromagnetic fields, $m_{sw} = 1/2A^*$ is the normalized effective mass of the wave packet.

Here, it is noted that the two equations are inter-correlated because Eq. (7a) contains term with the collective coordinates $x_i(t)$ of spin wave and Eq. (7b) contains term with the collective coordinates $X_\mu(t)$ of skyrmion. Particularly, the first and second terms in the right side of Eq. (7a) are associated with the effective electric and magnetic forces acting on the skyrmion respectively, the third term is the drag force, and the last term is the acting force imposed by the excited spin wave. Similarly, the motion of a single spin wave packet is also equivalent to the classical motion of a massive particle, which is determined by the effective forces acting on the wave packet expressed in the right side of Eq. (7b).

Certainly, it would be very useful if one can obtain a set of rigorous solutions to Eq. (7) in terms of motions of both skyrmion and spin wave packet. However, such a set of solutions seems to be challenging if not impossible, and therefore we may discuss a relatively simplified version of the motion behaviors. One case is to neglect the dissipative components which would be quite weak if the system under investigation can be optimized.

Given this assumption, one can update Eq. (7) for a simplified description of the motions of skyrmion and wave packets in the vector version, by setting damping factor $\alpha = 0$, yielding:

$$\rho M^{\mu\nu}\ddot{\mathbf{X}} = -(2\rho\omega\rho_{sw} - \delta_s)\mathbf{B}^0 \times \dot{\mathbf{X}} - \rho_{sw}\Omega_\pm \mathbf{b}\times\dot{\mathbf{x}} \quad (8a)$$

$$\Omega_\pm^2 m_{sw}\ddot{\mathbf{x}} = -\Omega_\pm \mathbf{b}\times\dot{\mathbf{x}} + (2\rho\omega - \delta_s)\mathbf{b}\times\dot{\mathbf{X}} \quad (8b)$$

where $\mathbf{B}^0$ and $\mathbf{b}$ are the equivalent fields acting on the skyrmion and spin wave packet, respectively. $\dot{\mathbf{X}}$ and $\ddot{\mathbf{X}}$ are the velocity and acceleration of the skyrmion, respectively, and $\ddot{\mathbf{x}}$ is the acceleration of spin wave packet. Eq. (8) represents a simplified version of Eq. (7) and the physics associated with the motions of skyrmion and spin wave packet can be seen more clearly from Eq. (8).

Subsequently, we present some qualitative discussion on these motions and the details of numerical calculations will be presented in Sec. III. As stated above, the motions of skyrmion and spin wave packet are inter-correlated. The spin wave driven motion of skyrmion can be discussed from two aspects. One is to check the scattering of spin wave packet from the skyrmion, and of course such scattering can be viewed as the counter-resistance of moving skyrmion against the spin wave as the driving force. The other is to check the skyrmion momentum imposed by the spin wave. Discussing the two aspects respectively allows us to understand the whole landscape of dynamics of the spin wave driven skyrmion motion.

*C. Scattering of spin wave packet*

Without losing generality, the scattering of the spin wave packet in the present problem can be discussed by assuming the approximately stationary skyrmion, as seen from Eq. (8b), given the much quicker motion of the spin wave packet than the skyrmion. The first term in the right side of Eq. (8b) is the effective Lorentz force acting on the spin wave packet, imposed by the skyrmion, which induces the transverse motion of the packet. This is the dominant term of the spin wave scattering. Moreover, by the definition $\Omega_\pm = 2\rho\omega \pm (-\delta_s)$, one sees that $\Omega$ has the same sign as the wave frequency $\omega$, noting that term $2\rho\omega$ gives the main contribution. In the other words, due to $|\delta_s| \ll 2\rho\omega$, the magnitude of $\Omega_\pm$ may change linearly with $\delta_s$, but the sign of $\Omega$ is determined by the handedness of the spin wave, i.e. positive $\Omega_+$ (negative $\Omega_-$) is obtained for the right- (left-) handed spin wave.

As a result, unless the net angular momentum $\delta_s$ is extremely large in magnitude, corresponding to the case of FM system, the spin wave scattering by the skyrmion in a FiM system would be hardly affected by $\delta_s$. The scattering would be likely similar to the case in an AFM system, as reported in recent literature. However, the skyrmion momentum imposed by the spin wave in a FiM system is rather different from those in FM and AFM systems, as revealed in the following section.

*D. Skyrmion momentum*

It is well known that for spin wave driven skyrmion motion in FM and AFM systems, the skyrmion motion is much slower than the propagation of spin wave packet. This fact applies to the present FiM system, and thus we first discuss the momentum onto the skyrmion imposed by the spin wave packet. Given the spin Berry phase term $L_B$ defined in Eq. (5), one may start from Eq. (7a) or Eq. (8a), to calculate the skyrmion momentum $P_\mu$ that is the derivative of $L_B$ with respect to the collective velocity defined by the collective coordinate $X_\mu$ [31]:

$$P_\mu = \partial L_B / \partial \dot{X}_\mu = \delta_s A_\mu^0 + \rho M^{\mu\nu} \dot{X}_\mu, \qquad (9)$$

The two terms constituting momentum $P_\mu$ are similar to those derived for FM lattice and AFM lattice, which are referred to FM term and AFM term, respectively. Notably, the skyrmion momentum in a FiM system depends on $\delta_s$ and this dependence does not exist in FM or AFM system. Obviously, the first term is linearly related to $\delta_s$ and symmetric about the $X$-axis, due to the mirror symmetry [13]. In the second term, parameter $\rho$ monotonously decreases with $\delta_s$, thus this second term is a decreasing function of $\delta_s$.

Consequently, parameter $\delta_s$ can be used as an effective control of the relative weight of the two momentum terms (i.e. their ratio), and eventually it becomes a critical parameter to determine the skyrmion dynamics. For the two limitations $\delta_s \to \infty$ and $\delta_s = 0$, skyrmion dynamics should be similar to those in FM system and AFM one, respectively. Importantly, the skyrmion Hall angle can be modulated via tuning $\delta_s$, which is very essential for future applications because the issue of Hall motion has been a highly concerned issue and detrimental for successful applications. Details of these behaviors will be presented below by the numerical calculations.

### III. Numerical analysis

The rigorous treatment presented above unveils quite a few of major characters of the skyrmion motion as driven by spin wave of circular polarization (right- or left-handedness). In order to check these predictions on one hand, and more importantly in order to apply this theory to a realistic system, one needs to comprehend our understanding of this theory by comparing the theoretical analysis with the numerical simulations based on the standard Heisenberg spin model using the LLG equation. The consistence of the proposed theory with the simulation results would be the criterion for the validity of the dynamic theory.

*A. Numerical calculations*

Based on Eq. (8), one notes that the motion of spin wave packet in connection with the skyrmion is similar to the motion of a classical particle under an effective magnetic field. However, the rigorous solutions to Eq. (8) are still challenging to obtain, and we turn to the numerical calculations using the fourth-order Runge-Kutta methods with the time step of $10^{-13}$ s. The numerical calculations are performed on a 400 nm × 400 nm system, in which the spin wave is excited at the left boundary and moves towards the skyrmion locating in the center of the system. As discussed previously, the skyrmion is pinned at this stage, considering the fact that the spin wave travels much faster than the skyrmion.

The velocity of spin wave can be calculated from the spin wave dispersion. Following the earlier work [44], we obtain the spin wave dispersion

$$\omega_{\pm} = \left[\pm\delta_s + \sqrt{\delta_s^2 + 4\rho(Ak^2 + K)}\right]/2\rho, \quad (10)$$

where $\omega_+$ ($\omega_-$) corresponds to the right- (left-) handed spin wave, $k$ is the wave vector of spin wave. Then, the speed of spin wave $v_{sw}$ is obtained from $v_{sw} = \partial\omega_{\pm}/\partial k$. Subsequently, the scattering of spin wave packet is calculated numerically by solving Eq. (8). Without loss of generality, we set the skyrmion topological number $Q = -1$, the skyrmion size to be 20 nm which corresponds to the case of $A = 15$ pJ/m, $D = 1.2$ mJ/m$^2$, and $K = 85$ kJ/m$^3$, and the spin wave frequency to be $\omega = 300$ GHz. These parameters are comparable with those as discussed in literature on compound GdFeCo [6,36].

B. Model-based simulations

We check and solidify our conclusion by micromagnetic simulations. For the present case, the model Hamiltonian on a generalized spin lattice may be given by

$$H = J\sum_i \mathbf{s}_i \cdot \mathbf{s}_{i+1} + D^*\sum_i (\mathbf{u}_{ij} \times \hat{z}) \cdot (\mathbf{s}_i \times \mathbf{s}_{i+1}) + K^*\sum_i (\mathbf{s}_i \cdot \hat{z})^2, \quad (11)$$

with the exchange coupling $J = Ad/4$ with the lattice constant $d$, the interfacial DMI magnitude $D^* = Dd^2/8$, the unit vector $\mathbf{u}_{ij}$ connecting the nearest neighbors, and the anisotropy constant $K^* = Kd^2d_z/2$ with the thickness $d_z$ [44]. The coupling parameters are the same as those in the numerical calculations. The dynamics of the skyrmion and spin wave packet is investigated by solving the LLG equation,

$$\frac{\partial \mathbf{s}_i}{\partial t} = -\gamma_i \mathbf{s}_i \times \mathbf{H}_{eff,i} + \alpha_i \mathbf{s}_i \times \frac{\partial \mathbf{s}_i}{\partial t}, \quad (12)$$

where $\mathbf{H}_{eff,i} = M_i^{-1}\partial H/\partial \mathbf{s}_i$ is the effective field with the magnetic moment $M_i$ at site $i$, the gyromagnetic ratio $\gamma_i = g_i\mu_B/\hbar$ with the g-factors [47] $g_1 = 2.2$ and $g_2 = 2$, and the damping constants are set to be $\alpha_1 = \alpha_2 = 0.004$.

For practical simulations based on the LLG equation, we start from a discrete lattice with the size of 400 nm × 400 nm × 7 nm and cell size of 2 nm × 2 nm × 7 nm, and set the time step to $10^{-13}$ s. The used magnetic moments $M_i$ for nine different cases are shown in Table 1 [36], and they correspond to nine different $\delta_s$. Here, the spin waves are excited by applying the AC magnetic field in the region of 70 nm < $x$ < 80 nm and 0 < $y$ < 400 nm, which propagate along the $x$-direction. Specifically, we generate right-/left-handed spin waves by applying AC magnetic field $\boldsymbol{h}_R/\boldsymbol{h}_L = h[\cos(\omega_h t)\mathbf{e}_x +/- \sin(\omega_h t)\mathbf{e}_y]$ with amplitude $h$ and the frequency $\omega_h = 300$ GHz [46]. The absorbing boundary conditions are used to eliminate the reflection of the spin waves at the boundary.

*C. Comparison between the numerical and simulated results*

Subsequently, we check the consistence of the numerical calculations based on Eq. (8) with the LLG simulations via the comparison between the numerical and simulated spin wave scattering. One notes that the driving force for skyrmion motion from the spin wave originates from the skyrmion-spin wave interaction, and it is intrinsic from the scattering of the spin wave by moving skyrmion. This scattering acts reversely on the skyrmion, driving its motion. Therefore, it is necessary to formulate the scattering process and then to discuss the underlying physics.

The Eq. (8)-based numerically calculated trajectories of the spin wave packet with right-handedness for various $\delta_s$ are shown in Fig. 1(a). It is clearly shown that the right-handed spin wave packet is scattered by the skyrmion towards negative $y$ direction. With the increase of $\delta_s$, the effective field of the skyrmion increases, which slightly enhances the scattering of the spin wave packet. Oppositely, the left-handed spin wave packet is deflected towards positive $y$ direction, as clearly shown in Fig. 1(b) where presents the scattering of the spin wave packet with left-handedness. Thus, the numerical calculations demonstrate that the scattering of spin wave is determined by the wave handedness and hardly affect by $\delta_s$, consistent with the above analysis.

This property is also confirmed in our micromagnetic simulations. Fig. 2 presents the Bragg intensity from Fourier transformation of the spatiotemporal oscillation of simulated $n_x$, which describes the spin wave intensity. For the right-handed spin wave, considerable Bragg intensity is observed in the vector region with negative $k_y$ as depicted with red circles in Fig. 2(a)-2(c), confirming that the spin waves are scattered towards negative $y$ direction by the skyrmion. On the other hand, the left-handed spin waves are deflected towards positive $y$ direction, resulting in the considerable Bragg intensity in the positive $k_y$ region, as shown in Fig. 2(d)-2(f). Undoubtedly, the simulations are well consistent with the numerical calculations, confirming the validity of theoretical analysis above.

While parameter $\delta_s$ only weakly affects the spin wave scattering, its effect on the spin wave intensity is relatively remarkable and depends on the spin wave handedness, as shown in Fig. 2. It is clearly shown that for the right-handed spin waves, the wave magnitude increases with increasing $\delta_s$. This effect is attributed to the resonant frequency of spin wave in FiM systems, which reads

$$\omega_{\pm}^c = \left(\sqrt{(J+K)^2 \delta_s^2 + (8JK+4K^2)s_1 s_2} \pm \delta_s (J+K)\right)/2s_1 s_2. \tag{13}$$

Obviously, as the frequency of excitation field $\omega_h$ gets close to the resonant frequency, high-intensity spin wave is excited. Here, $\omega_+^c$ is estimated to be ~22.5 GHz and further increases with increasing $\delta_s$, resulting in the enhancement of spin wave intensity. Reversely, for the left-handed spin wave, $\omega_-^c$ decreases and gets far away from $\omega_h$ with the increase of $\delta_s$, resulting in the suppression of the spin wave intensity, as shown in Figs. 2(d)-2(f).

As a matter of fact, the scattering of spin wave by skyrmion in FM [30] and AFM [13] systems has been respectively discussed in earlier works. In these systems, the scattering direction or scattering cross-section determines the momentum transfer from the spin wave to the skyrmion, which in turn modulates the skyrmion motion, similar to the course of two particles elastic collision. However, in a FiM system, the case becomes more complex because the relative weight of the two skyrmion momentum terms (i.e. their ratio) significantly depends on $\delta_s$, as revealed in section II. D. Thus, it is reasonably expected that $\delta_s$ is a critical parameter to determine the skyrmion dynamics including Hall motion, although it hardly affects the scattering of injected spin wave.

*D. Effect of $\delta_s$ on skyrmion Hall motion*

The consistence of the micromagnetic simulations with the numerical calculations allows one to uncover clearly the effect of $\delta_s$ on the skyrmion dynamics using simulations, noting that a rigorous derivation of the skyrmion velocity is quite challenging.

As predicted in Section II. D, $\delta_s$ can effectively modulate the ratio between the FM momentum term and the AFM term, which in turn tunes the skyrmion Hall motion. This prediction is verified in our micromagnetic simulations, and interesting dynamic behaviors of the skyrmion are revealed. Fig. 3(a) presents the trajectory of the skyrmion for various $\delta_s$ driven by the right-handed spin waves excited by magnetic field with $h$ = 600 mT. One notes that the FM momentum term is completely suppressed for $\delta_s$ = 0, and the skyrmion dynamics in ferrimagnets are controlled by the AFM term. Thus, strong Hall motion is observed at $\delta_s$ = 0, the same as in AFM system [13] due to the fact that dynamics of a magnetic texture mainly depends on the net angular momentum. In this case, the ferrimagnetic skyrmion propagates away from the spin wave source with a strong transverse motion along the +$y$ direction (black squares in Fig. 3(a)).

With the increase of positive $\delta_s$, the FM term enhances significantly with the suppression of the AFM term. Moreover, the two terms contribute to the transverse motion along the +$y$ direction, while compete with each other to the longitudinal motion. As a result, the Hall angle increases gradually, and the skyrmion moves toward the spin wave source for $\delta_s$ > 0.3 (×10$^{-7}$ Js/m$^3$). When $\delta_s$ increases to ~ 1.24 where the FM term dominates, the skyrmion dynamics approaches to that of ferromagnetic skyrmion with topological number $Q$ = −1 (depicted by blue dashed arrow). On the other hand, for negative $\delta_s$, the two terms compete with each other to both the longitudinal and transverse motion components, contributing to the decrease of the skyrmion speed with the increasing |$\delta_s$|. Similarly, with the increase of |$\delta_s$|, the skyrmion motion gradually approaches to that of ferromagnetic skyrmion with opposite charge $Q$ = +1. The skyrmion Hall angle as a function of $\delta_s$ is summarized in Fig. 4(a), which clearly demonstrates that the Hall angle increases with the increasing $\delta_s$.

From the symmetry of the two sublattice system, opposite Hall angles are expected for two ferrimagnetic skyrmions with opposite $\delta_s$ driven by the left-handed and right-handed spin waves, respectively, which has been confirmed in our simulations. In Fig. 3(b), we present the skyrmion trajectory for various $\delta_s$ driven by the left-handed spin waves excited by magnetic field with $h$ = 1200mT. It is clearly shown that the trajectories of two skyrmions with opposite $\delta_s$ driven by the right-handed and left-handed spin waves respectively ($\delta_s$ = 0.62, for example, blue empty triangles in Fig. 3(a) and solid triangles in Fig. 3(b)) are symmetric around the *x*-axis, demonstrating the opposite Hall angles for the two cases as shown in Fig. 4(a). It is clearly shown that the dependence of Hall angle on $|\delta_s|$ driven by the left-handed spin waves is similar to the right-handed spin-wave-driven one.

So far, we have demonstrated theoretically and numerically that skyrmion Hall motion in ferrimagnet highly depends on $\delta_s$ via tuning the ratio between two momentums terms acting on the skyrmion, allowing one to select suitable materials to control the skyrmion motion better. Furthermore, it is worth noting that the amplitude of spin wave is also related to $\delta_s$, while it hardly affects the skyrmion Hall angle.

At last, we check the effect of spin wave frequency $\omega_h$ on the skyrmion Hall motion. As a matter of fact, earlier work has reported that in ferromagnets and antiferromagnets, the increase of spin wave frequency at low wave number enhances the wave group velocity [13]. Thus, the effective Lorenz force is also increased, resulting in an enhanced skyrmion Hall motion. This property is also available in ferrimagnets. In Fig. 4(b), we presents the simulated Hall angle as a function of $\omega_h$ for various $\delta_s$ at the wave number ~ 0.2 nm$^{-1}$, which clearly demonstrates the slight increase of the angle with the increasing $\omega_h$ regardless of the $\delta_s$ value.

*E. Connection with experiments*

Actually, one may estimate the speed of the skyrmion to provide more guidance for experiments, noting that the parameters considered in this work are comparable with those in GdFeCo. Specifically, a speed of ~320 m/s is obtained for $\delta_s$ = 0.31 (10$^{-7}$Js/m$^3$), comparable with that driven by electric current [10]. Thus, spin wave is revealed to be an efficient stimulus in driving skyrmions in ferrimagnets, which will be highly preferred in future low-power-consuming device design.

More importantly, $\delta_s$ has been proven to be a core control parameter in modulating the skyrmion Hall motion, allowing one to better control the skyrmion dynamics through tuning $\delta_s$ value in ferrimagnets. For example, suitable $\delta_s$ can be adjusted by ion doping to suppress the skyrmion Hall motion, which is of great importance for future applications considering the detrimental effect of Hall motion on data propagation. Of course, the prediction of spin-wave driven skyrmion dynamics in ferrimagnet given in the work deserves to be checked in further experiments.

## IV. Conclusion

In conclusion, we have studied theoretically and numerically the skyrmion dynamics in ferrimagnets driven by circularly polarized spin waves. The spin wave scattering highly depends on its chirality, and the scattering direction hardly be affect by the net angle momentum $\delta_s$. Interestingly, $\delta_s$ determines the ratio between the two spin-wave induced momentum terms acting on the skyrmion, which in turn significantly affects the skyrmion Hall motion. The dependence of the Hall angle on $\delta_s$ and spin wave frequency is unveiled by numerical simulations, which demonstrate antiferromagnetic dynamics at $\delta_s = 0$ and ferromagnetic-like dynamics for large $\delta_s$. More importantly, the gradual transition in Hall angle on $\delta_s$ allowing one to select suitable materials to better control the skyrmion motion, which is very meaningful for future spintronic applications.

**Acknowledgment**

The work is supported by the Natural Science Foundation of China (Grants No. 51971096, No. 92163210, and No. 51721001), and the Natural Science Foundation of Guangdong Province (Grant No. 2019A1515011028 and No. 2022A1515011727).

Table 1. Parameters used in the numerical simulations.

| Index | 1 | 2 | 3 | 4 | 5 | 6 | 7 | 8 | 9 |
|---|---|---|---|---|---|---|---|---|---|
| $M_1$(kA/m) | 1120 | 1115 | 1110 | 1105 | 1100 | 1095 | 1090 | 1085 | 1080 |
| $M_2$(kA/m) | 1040 | 1030 | 1020 | 1010 | 1000 | 990 | 980 | 970 | 960 |
| $\delta_s$(×$10^{-7}$Js/m$^3$) | −1.24 | −0.93 | −0.62 | −0.31 | 0 | 0.31 | 0.62 | 0.93 | 1.24 |

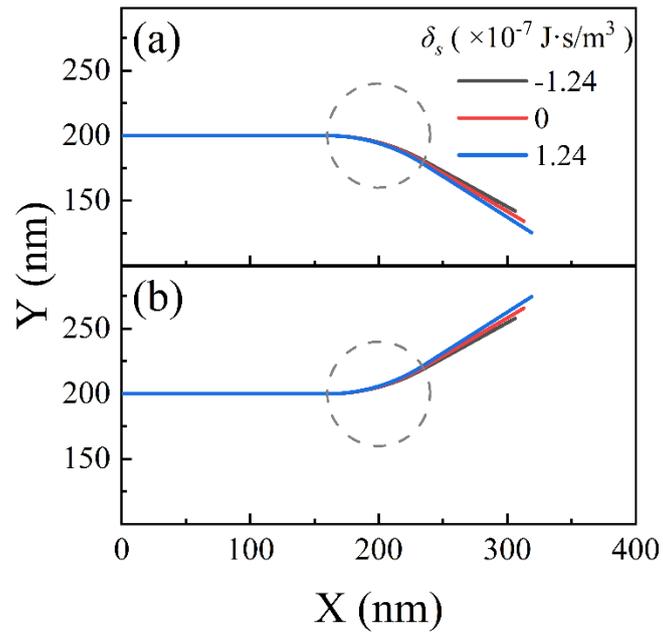

Fig. 1 Skew scattering of (a) right-handed and (b) left-handed spin wave packet across the skyrmion (dashed circle).

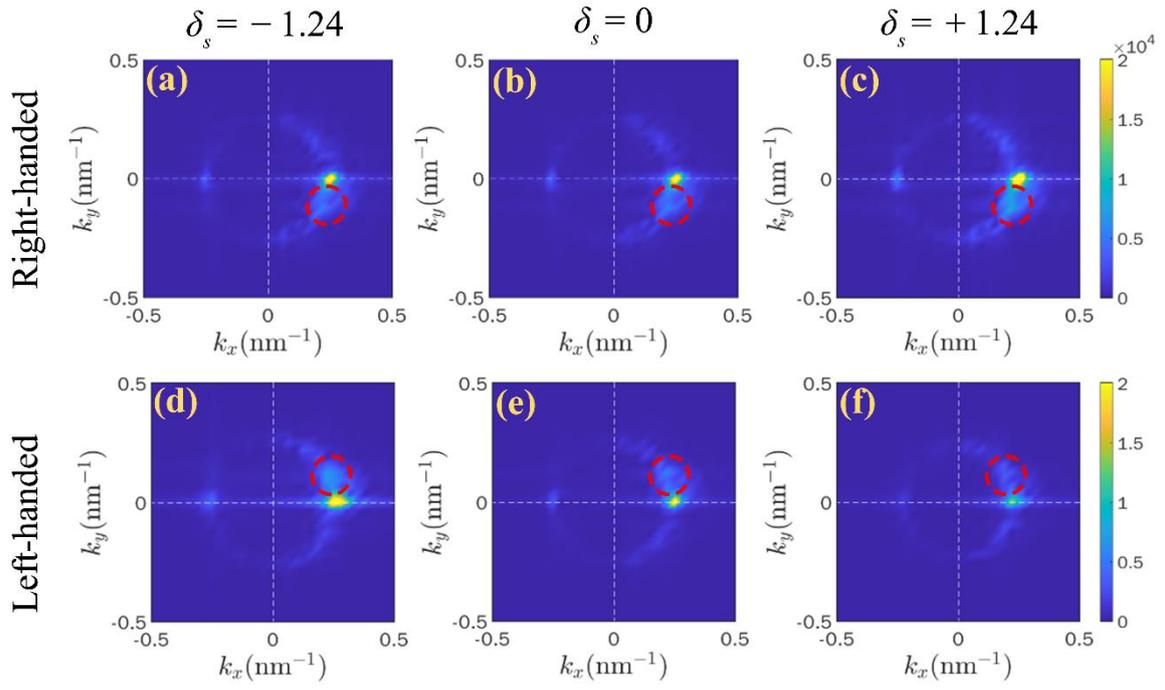

Fig. 2. Spatial fast Fourier transition spectrum for $\omega_h$ = 300GHz for various $\delta_s$ in unit of $10^{-7}$ Js/m$^3$ driven by right-handed spin waves (a, b, c) and by left-handed spin waves (d, e, f). Here, the fast Fourier transition analysis is implemented over the whole plane.

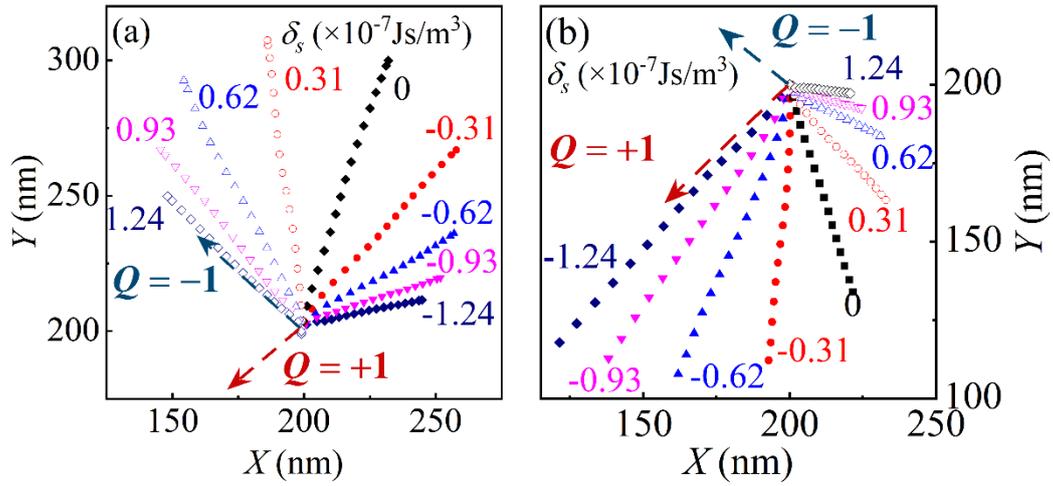

Fig. 3. The trajectory of skyrmion driven by (a) the right-handed spin waves, and (a) the left-handed spin waves for various $\delta_s$.

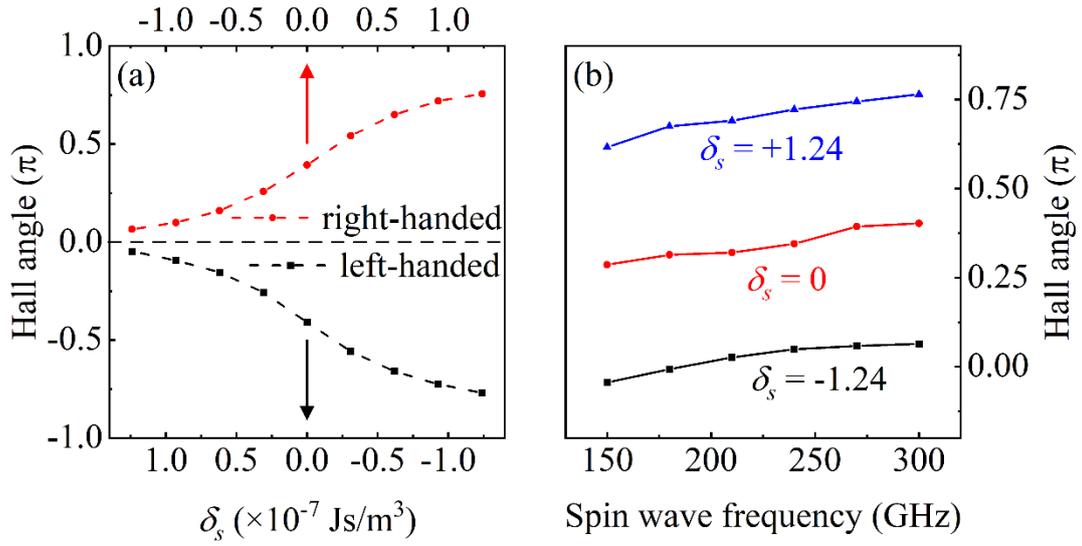

FIG. 4. The simulated Hall angle (a) as a function of $\delta_s$ driven by the right-handed and left-handed spin waves for $\omega_h = 300$ GHz, and (b) as a function of $\omega_h$ driven by the right-handed spin waves for various $\delta_s$.